\documentstyle[preprint,psfig,epsf,aps]{revtex}
\begin{document}
\draft
\tightenlines
\title{\bf{Roughening of a growing 
surface on a crystal with correlated disorder:
influence of nonlinearity} }
\author{ Sutapa Mukherji}
\address{Institut f\"ur Theoretische Physik,
Universit\"at zu K\"oln, Z\"ulpicher Str. 77, D 50937 K\"oln, Germany}
\maketitle
\begin{abstract} 
We study the  growth of a crystal  in  presence of a  correlated 
disorder on the substrate. Using functional renormalization group, 
 we show, for a long range disorder correlation, an initial 
decay of the KPZ type 
nonlinearity, though over a 
large length scale the behavior can be governed by the  nonlinearity.
\end{abstract}
\section{INTRODUCTION}
The equilibrium shape of a crystal surface undergoes a roughening 
transition from a rough high temperature phase to a smooth low 
temperature phase as the temperature is decreased \cite{van,noz,chui}.
Above the roughening transition $T_r$, the height fluctuations grow 
logarithmically with the dimension of 
the system $L$  and below the transition the height is smooth and  
 is  independent of $L$.
 The non-equilibrium counterpart, the  growth mechanism 
of such crystals,  provides more insights about the 
roughening transition \cite{noz}. 
It is found that for $T>T_r$ the growth is non-activated in nature
whereas for $T<T_r$, the growth is essentially by nucleation of  droplets
and the growth velocity is exponentially slow in the inverse of the force.
For an infinitesimal force $F$, the mobility which is the ratio of the 
growth velocity and $F$ 
vanishes with a  jump from a finite value at the transition.
With a finite force the transition is blurred. The flat phase is 
destroyed over  large length scale and eventually becomes rough.
In this case a continuous 
decrease of the mobility is observed as the temperature is lowered
\cite{noz}.

An interesting development in this direction is the inclusion
of the disorder of the underlying substrate \cite{tsai1}. 
The morphology of the growing
surface  shows  a new transition which is the  super
roughening transition at temperature $T_{sr}$.
For $T>T_{sr}$, the height fluctuation is the same as  the high 
temperature phase of the usual roughening transition but for 
 $T<T_{sr}$ the surface is rougher than the thermal phase.
There is a controversy regarding the roughness
 of this super rough phase
\cite{cardy,korsh,cule} though a recent numerical treatment \cite{shapir} 
favors the size dependence of the roughness as $(LogL)^2$.
A dynamical renormalization group
treatment \cite{tsai1}
  shows that for $T>T_{sr}$ the effect of the 
disorder essentially vanishes
over large length scales and the  scaling properties of the surface
are the same as that of a surface growing on a pure substrate, in the rough 
phase. The linear response mobility in this super roughening transition
 vanishes
continuously, unlike a jump  in the case of the  roughening 
transition, as 
$T_{sr}$ is approached from the high temperature phase.
 For $T<T_{sr}$
there is a temperature dependent dynamical exponent and a  
nonlinear response of the system to an external force.
The connection between these roughening and super 
roughening transitions has been elucidated furthermore 
by a renormalization group treatment of the growth on  a substrate with 
a correlated disorder \cite{stef}. 
The roughening transition turns into a super roughening
 transition
as the correlation of the disorder  decays sufficiently fast. By tuning the 
power law of the long range correlation of the disorder it is possible to 
go from the pure to the short range  disorder limit.

Such growth problems for crystals need to take into account 
the periodicity
perpendicular to the crystal surface and is usually 
incorporated in the growth equation by a periodic pinning potential.
 The low temperature, flat phase in the roughening transition is 
due to the relevance of this pinning potential.

The nonequilibrium growth problem is complicated. One has to take 
into account a relevant nonlinear term that appears due to the 
lateral growth of an oblique surface \cite{kpz}. 
One of the major consequences of this Kardar-Parisi-Zhang (KPZ) 
nonlinearity is a 
power law (not logarithmic)  growth of height fluctuations. In the 
context of the nonequilibrium situation of a 
crystal surface, it has been argued that with a finite force such 
a nonlinear term is generated by the interplay  of the force and this  
pinning potential \cite{rost}.
 The asymptotic behavior of the surface might then be determined by the
 nonlinearity which can destroy the roughening transition.
A rough phase can appear even at low temperature.
If one starts with a nonlinear term, the lateral growth and 
the pinning potential combinedly lead to a phase factor in the pinning 
potential \cite{tsai2}. This phase factor is also 
renormalized as one looks at the system over  larger length scales.
The renormalization of the phase factor makes the identification of 
the phases a nontrivial problem. 

The disorder substrate case is also not well understood.
Previous analysis in this direction with a short range correlation 
of the disorder of the lattice showed that in the presence of the
nonlinearity a small 
driving force is relevant and 
the asymptotic properties are essentially
governed by the nonlinear term since the finite velocity leads to a 
smearing of the pinning potential \cite{tsai2}.
 A numerical investigation on the other hand  showed
a generation of a quenched random mobility \cite{krug} and hence a 
new universality class. The latter has 
been attributed essentially to the effect  of a finite 
lattice cutoff.

In this paper, we address the question of a large scale description of 
the dynamics of a growing surface in the 
 presence of a  disorder on the underlying 
lattice and a KPZ type nonlinearity discussed before.
We consider a very general form of the 
disorder correlation. In view of the conflicting scenario mentioned above 
our
analysis is based on a renormalization group technique that involves
a finite lattice cutoff. 
A standard way of describing such a growth phenomena
 is to  start from the equation of motion for the height 
$\phi({\bf r},t)$ at time $t$ and at coordinate  ${\bf r}$  in the two
dimensional plane. As mentioned before 
the lattice structure orthogonal to the 
crystal surface is 
respected through the inclusion of a periodic potential.
 The equation of motion we need to study 
is therefore the  usual Sine-Gordon equation 
\cite{tsai1,stef,rost,tsai2,krug} subjected to a
constant driving force $F$
\begin{equation}
 m^{-1}\frac{\partial \phi({\bf r},t)}{\partial t}=
K \nabla^2 \phi({\bf r},t)-V\ \sin[\phi({\bf r},t)+d({\bf r})]+
{\overline{\lambda}}/2(\nabla \phi)^2+F+R({\bf r},t) \label{eq2}
  \end{equation} 
where $K$ denotes the stiffness of the surface, $m$ is the microscopic
mobility,
 $V$ is the strength of the pinning potential. Here ${\overline{\lambda}}$
 represents the 
strength of the nonlinearity that allows a lateral growth of the surface
and $R$ represents the thermal noise at temperature $T$ 
with a short range correlation
given by 
\begin{equation}
  \langle R({\bf r},t)R({\bf r}',t')\rangle=
(2T/m)\ \delta({\bf r}-{\bf r}')\delta(t-t')\label{eq3}
\end{equation}
In the presence of a random substrate we expect that the minima 
of the  periodic 
potential will be randomly shifted and this is incorporated in this equation
through a quenched random variable $d({\bf r})$.
We consider here a general form of the correlation 
 associated with this random phase shift $d({\bf r})$,
as $g_0\overline{ e^{i(d({\bf r})-d({\bf r}'))}}=\gamma(r-r')$, where 
$g_0=\frac{1}{2}m^2 V^2$ and the
overbar denotes  average over the disorder. At this point we 
 do not specify the functional form of $\gamma(r-r')$ but later we 
concentrate on $\gamma(r)\sim r^{-2\alpha}$. We obtain back the   
short range disorder case  for large $\alpha$ and the 
perfect crystal for $\alpha=0$. 

Various limiting forms of this equations  which have been studied are as 
follows.
(i) The most well studied limit is $V=0$ \cite{halpin}.
(ii) The pure equilibrium growth that shows a roughening transition is 
obtained with ${\overline{\lambda}}=F=d=0$ \cite{chui,noz}.
The width of the growing surface is conventionally described by the 
following scaling $w(L,t)=L^{\chi}f(t/L^z)$,
where $\chi$ is the roughening exponent and $z$ is the 
dynamical exponent.
Above the roughening transition, the width scales as 
$w(L,t)\sim \ln[Lf(t/L^z)]$,
which implies $\chi=0$.  In the low temperature flat phase $w(L,t)$
 is independent of $L$.
(iii) The nonequilibrium situation of this pure problem  is described by 
equation \ref{eq2} with $d=0$, with or without $F$ \cite{rost}.
(iv) The short range correlated disorder on the substrate follows from 
$\gamma(r)=g_0\delta(r)$ \cite{tsai1}. This shows the continuous super
roughening transition.
(v) A long range correlated disorder with $\gamma(r)\sim \ r^{-2\alpha}$
and ${\overline{\lambda}}=0$ shows the 
interpolation between the roughening and 
super roughening transition \cite{stef}.
Another interesting limit, $V={\overline{\lambda}}=0$ 
and a quenched noise 
$R(\phi,{\bf r})$ was 
used to study the dynamics of driven interface
in a disordered medium \cite{natter}. 

In a renormalization group (RG) approach the system is looked 
at longer length scales by integrating out the effects of small scale 
fluctuations. When the system is rescaled to the original scale, the effect 
of the small scale fluctuations goes into the renormalization of the 
various parameters of the problem. The effective coupling constant
observed at a certain length scale is then given by the RG recursion 
relations. In this approach, in all the above cases except 
(ii) with $F\neq 0$, $\lambda$ has been found not to be renormalized.
The important new feature that arises in our case 
is the renormalization of the nonlinearity due to the 
nonlocal property of the correlator. We obtain a surprising result 
that this nonlinearity
decays initially with the length scale for $F=0$. However the 
 possible generation of a force and its
relevance can cause a truncation of this  decay of the
 nonlinearity and the growth asymptotically becomes KPZ like.
In order to treat the long range correlation in general, we use the functional
renormalization group (FRG) approach where the renormalization of the 
correlation 
function $\gamma(r)$ and its effect on the other parameters are studied.

The paper is organized as follows. Section II is devoted to the description
of the effective generating functional. In section III we derive the 
functional renormalization group flow equation with necessary diagrams. 
At this point we might add that our approach is simpler than the dimensional 
regularization approach of Ref \cite{tsai2}. Details of the 
calculations are presented in the appendices. Section IV
is devoted to the discussion of the asymptotic behavior  of the system
and connection with previous  predictions. In section V, we summarize 
our results.

\section{GENERATING FUNCTIONAL}

The growth on a crystal substrate with disorder is described by the phase 
disordered Sine-Gordon model with a KPZ nonlinearity as described in equation
\ref{eq2}.
We use the Martin Siggia Rose formalism \cite{martin}
 that requires
a response field $\tilde \phi({\bf r},t)$.
Averaging over the disorder \cite{comm} yields 
the  generating functional
$Z=\int {\cal D}\tilde \phi {\cal D}\phi\  \exp[{\cal A}]$. 
Here ${\cal A}={\cal A}_0^{(0)}+{\cal A}_0^{(d)}$ 
is the effective action with free 
and the disorder part of the action  given respectively by
\begin{mathletters}
\begin{eqnarray}
&&{\cal A}_0^{(0)}= \int dt d{\bf r}\{\frac{1}{2} {\it \vartheta}_0 
{\tilde \phi}_0({\bf r},t)^2-{\tilde \phi}_0({\bf r},t)[\mu_0^{-1}
\dot\phi_0({\bf r},t)-\kappa_0 
\nabla^2\phi_0({\bf r},t)-\frac{\lambda_0}{2}(\nabla \phi_0({\bf r},t))^2]
\nonumber\\
&&+{\tilde J}_0({\bf r},t)
{\tilde \phi}_0({\bf r},t)\}\label{eq8a}\\
&&{\cal A}_0^{(d)}=\int dt dt' d{\bf r}d{\bf r'}
 \frac{1}{2}\gamma_0({\bf r}-{\bf r'})
{\tilde \phi}_0({\bf r},t){\tilde \phi}_0({\bf r'},t')cos[\phi_0({\bf r},t)-
\phi_0({\bf r'},t')]\label{eq8}
  \end{eqnarray}
\end{mathletters}
where $\phi_0=\phi$, $\vartheta_0=2mT$, $\mu_0=1$, $\kappa_0=mK$, 
$\lambda_0={\overline{\lambda}} m$ 
and ${\tilde J}_0=mF$ are the bare quantities.
The Gaussian part of the action (Eq. \ref{eq8a} with $\lambda_0=0$ )
gives rise to the following response and 
the correlation functions\cite{tsai1} in the momentum and frequency 
representation,
\begin{eqnarray}
 \langle \phi({\bf q},\omega)\tilde \phi({\bf q'},\omega')\rangle =
R(q,\omega)\delta({\bf q}+{\bf q}')\delta(\omega+\omega'),\nonumber\\
\langle \phi({\bf q},\omega) \phi({\bf q'},\omega')\rangle =
C(q,\omega)\delta({\bf q}+{\bf q}')\delta(\omega+\omega'),
\label{eq9}
\end{eqnarray}
where, 
\begin{eqnarray}
R(q,\omega)=\frac{\mu}{-i\omega+\mu\kappa q^2},\nonumber\\
C(q,\omega)=\frac{\vartheta \mu^2}{\omega^2+\mu^2 \kappa^2 q^4}\label{eq10}
\end{eqnarray}
(suppressing  subscript 0).
In the momentum and time representation 
\begin{eqnarray}
\langle \phi({\bf q},t)\tilde \phi(-{\bf q},t')\rangle =\theta(t-t')\mu
e^{-\mu\kappa q^2 (t-t')},\nonumber\\
\langle \phi({\bf q},t) \phi(-{\bf q},t')\rangle =\frac{\mu\vartheta}
{2\kappa q^2}e^{-\mu\kappa q^2 \mid t-t'\mid},\label{eq10a}
\end{eqnarray}
 where $\theta(t)=1$ if $t>0$ and zero otherwise. In real space 
\begin{equation}
R({\bf r},t)=\frac{\theta(t>0)}{4\pi\kappa t} e^{-r^2/(4\mu\kappa t)}.
\end{equation}
The correlation function has a divergence due to long wave length and
for our purpose we  define a difference correlation with a suitable 
ultraviolet regularization introduced by a cutoff $\Lambda$ as following,
\begin{equation}
Y(\mid {\bf r}-{\bf r}'\mid,t-t')=
\frac{1}{2}\langle [\phi({\bf r},t)-\phi({\bf r'},t')]^2\rangle=
\frac{\mu\vartheta}{4\pi\kappa}
\int_{\mid {\bf k}\mid\leq\Lambda} \frac
{dk}{k} 
[1-J_0(k\mid {\bf r}-{\bf r}'\mid)
e^{-\mu\kappa k^2 \mid t-t'\mid}]
\end{equation}
Our analysis involves this difference correlation. 
The correlation and the response functions are also connected by the 
Fluctuation Dissipation theorem (FDT) given as
\begin{equation}
\theta(t>0) \partial_t C({\bf k},t)=-\frac{\mu \vartheta}{2}R({\bf k},t)
\label{eq10b}
\end{equation}
An important aspect of our system is the breaking of the FDT. This
violation of FDT has previously been argued in the dynamics of 
random phase sine Gordon model \cite{culey}.
The diagrams corresponding to the correlation and 
response functions are presented in Fig 1 along with the 
 other two vertices 
$\tilde \phi \tilde \phi' cos(\phi -\phi')$ and $\tilde \phi
(\nabla \phi)^2$, where ${\tilde {\phi'}}$ and $\phi'$ denote the fields
at $({\bf r}',t')$.  These are the basic diagrams needed for the subsequent 
perturbative FRG.
\vbox{
\begin{center}
\psfig{file=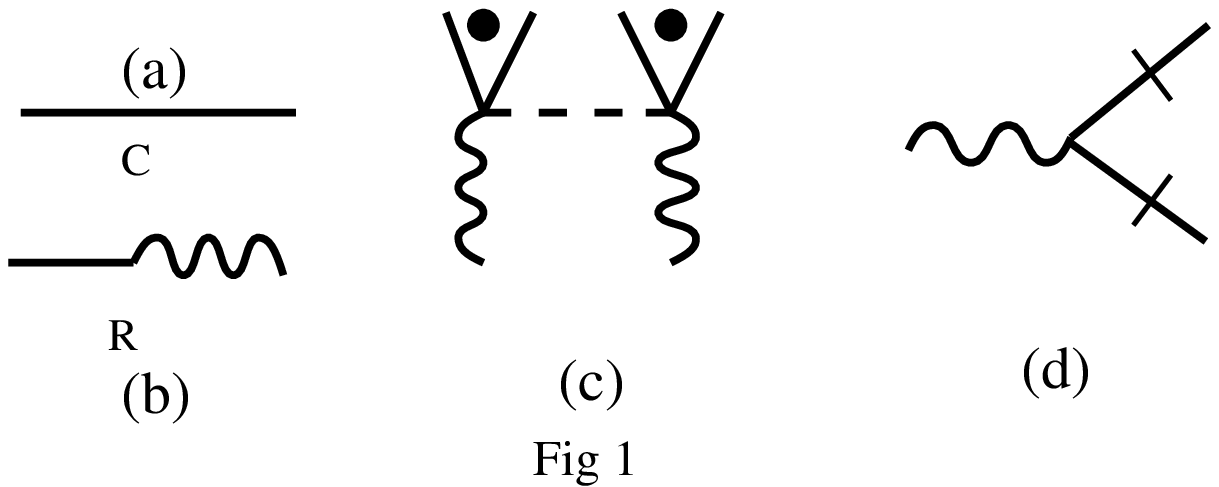,width=6in,angle=0}
{Fig 1. Diagrammatic representations of (a) correlation function,
(b) response function, (c) ${\tilde \phi({\bf r},t)}{\tilde \phi({\bf r}',t')}
cos[\phi({\bf r},t)-\phi({\bf r}',t')]$ 
where dots represent many $\phi$ lines, (d)
${\tilde \phi({\bf r},t)}(\nabla\phi)^2$}
\end{center}}

\section{Renormalization} 
In the following we discuss the renormalization scheme to obtain the 
physics at large distances, which, in the momentum space, corresponds to
small ${\bf k}$ \cite{shukla}. We  
 use the description of the fields as sum of the 
fast and slow modes
defined as, 
\begin{eqnarray}
\phi_s=\phi(k)\ \  {\rm for}\ \  0<k<\Lambda,\nonumber\\
\phi_f=\phi(k)\ \ {\rm for}\ \  \Lambda<k<\Lambda+\delta\Lambda.
\end{eqnarray} 
 This separation 
of fast and slow modes is used to average out the fast fourier modes 
or the short wavelength details of the problem.
 This short wavelength
 properties are incorporated by the appropriate 
renormalization of the parameters in the 
effective action which describes the large length scale properties.
By separation of the fast and slow Fourier modes,
in general, one arrives at the action,
\begin{eqnarray}
Z=&&\int \prod_{0 \leq k<\Lambda} {\cal D}\phi{\cal D}{\tilde {\phi}}
e^{{\cal A}_0(\phi_s,{\tilde {\phi_s}})} \int
\prod_{\Lambda \leq k<\Lambda+\delta \Lambda} {\cal D}\phi{\cal D}
{\tilde {\phi}} 
e^{{\cal A}_0(\phi_f,{\tilde {\phi_f}})}
e^{{\cal A}_I(\phi_s,{\tilde {\phi_s}},
\phi_f,{\tilde {\phi_f}})}\nonumber\\
=&&\int {\cal D}\phi_s {\cal D}{\tilde {\phi}}_s
 e^{{\cal A}'(\phi_s,{\tilde {\phi}}_s)},
\end{eqnarray}
where ${\cal A}_0$ is the free action and ${\cal A}_I$ is 
the interaction 
part that contains both fast and slow modes.
Here 
\begin{equation}
e^{{\cal A}'(\phi_s,
{\tilde {\phi}}_s)}=e^{{\cal A}_0(\phi_s,{\tilde {\phi_s}})}
\langle e^{{\cal A}_I(\phi_s,{\tilde {\phi_s}},
\phi_f,{\tilde {\phi_f}})}\rangle_{0>},
\end{equation}
where  $<\ >_{0>}$ denotes average with respect to fast modes of 
the free action.
The next step  is a cumulant expansion to clearly identify the 
contribution from averaging of the fast modes, as follows
\begin{equation}
e^{{\cal A}'(\phi_s)}=e^{{\cal A}_0+\delta {\cal A}},
\end{equation}
where \[\delta {\cal A}=\langle {\cal A}_I \rangle+
(\langle {\cal A}_I^2\rangle-
\langle {\cal A}_I \rangle^2)/2+ ...\]
 The system in
its original length scale is retrieved back again by  rescaling 
the fields. Rescaling implies that 
under the transformation $x\rightarrow bx$, 
$t\rightarrow b^z t$, $\phi\rightarrow b^{\chi}\phi$ and 
${\tilde \phi}\rightarrow b^{\tilde\chi}{\tilde \phi}$. 
Around the Gaussian fixed point we have 
$z=2$ and  in a $2$-d system the rescaling leads to 
${\tilde \phi}\rightarrow b^{-2}{\tilde\phi}$ with  
$\phi$  remaining invariant.
\vbox{\begin{center}
\psfig{file=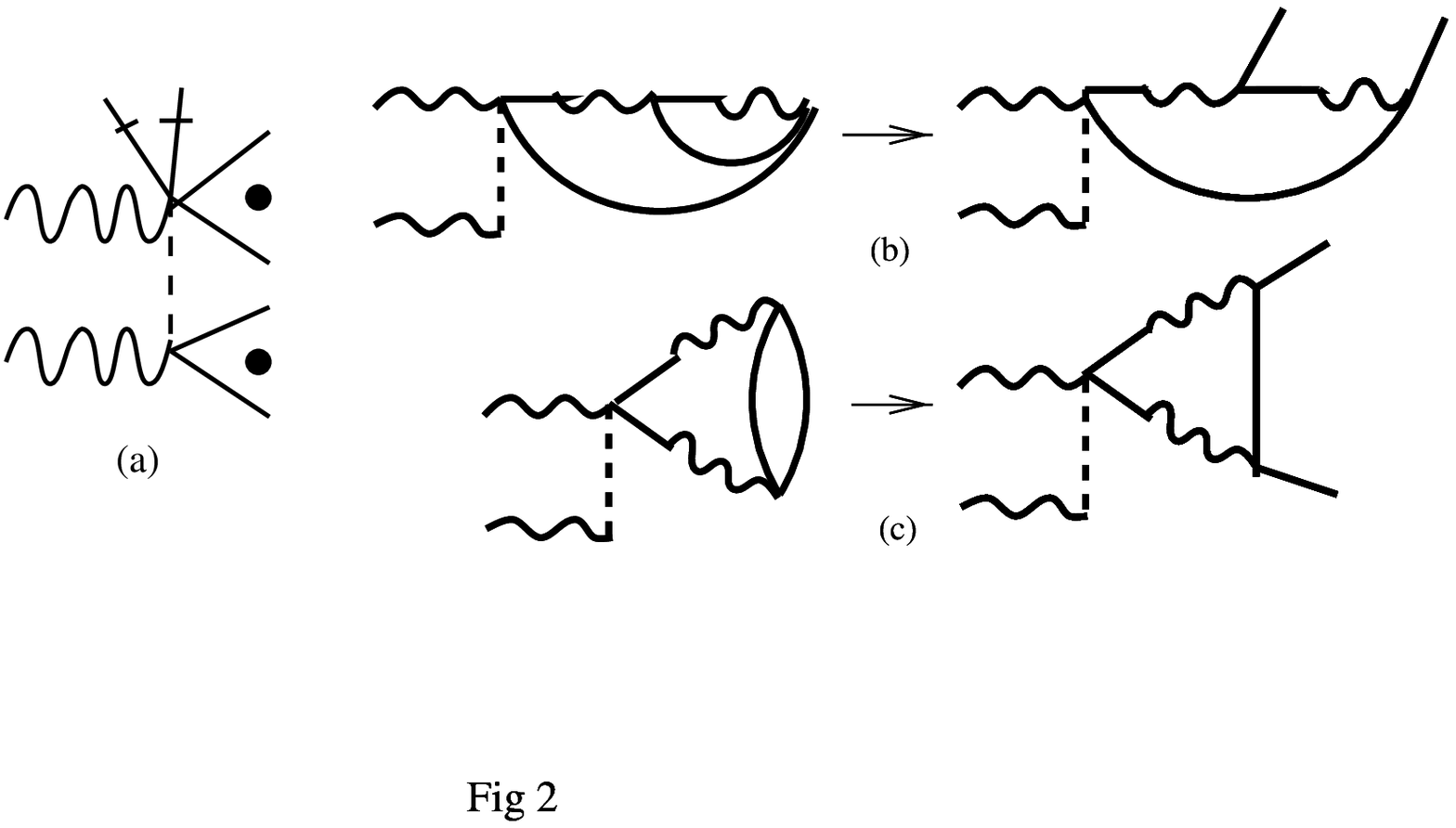,width=6in,angle=0}
{Fig. 2 (a) New vertex ${\tilde \phi({\bf r},t)}{\tilde \phi({\bf r}',t')}
(\nabla\phi({\bf r},t))^2 cos[\phi({\bf r},t)-\phi({\bf r}',t')]$.  
Hatched line represents
$\nabla\phi({\bf r},t)$, 
(b) and (c) Origin of this new vertex. Figures on the 
left side of the arrows show the two loop diagrams. The
 new vertex is generated by snipping off one internal line of the
two loop diagram. Diagrams found by this procedure are shown on the 
right side of the arrows.}
\end{center}}

The terms that contribute to the renormalization of various vertices 
have  diagrammatic representations that are  presented as we proceed.
In the following we consider  one loop diagrams. In fact in this approach 
since each internal line in the diagram is within the 
shell $\Lambda+\delta \Lambda$ it is  sufficient to consider the 
diagrams with only one loop \cite{wegner}.
 The effect of the higher loops in the field
theoretic approach \cite{tsai2} 
are taken care of in our approach by the new vertex 
generated under renormalization.
This  feeds back to the renormalization 
of the original vertex we start with.

The nonlinear term causes  a renormalization of the tension 
$\kappa$ and the 
temperature $\vartheta$. We shall not elaborate on this part since they are
very well documented in previous work \cite{kpz}. 
The mobility $\mu$, as well as 
$\vartheta$ and $\kappa$ are renormalized by the pinning potential. 
For local
correlator this can be found in references \cite{gs,tsai1}. 
The nonlocality of the correlator 
requires a more general treatment presented in  the appendix. 
There is a possible new renormalization of the 
pinning potential due to the nonlinearity.
A straightforward diagrammatic expansion would produce two loop diagrams
\cite{tsai2} as shown on the left sides of Fig 2b and Fig2c.To include 
the effect of these in our RG scheme, we have to involve new vertices whose
one loop contribution would be equivalent to the effect of original diagrams. 
This new vertex shown in Fig 2a is generated by the 
combination of $\lambda$ and $\gamma(r)$ and will be denoted in the following 
as $\gamma_1(r)$. The explicit form of this vertex is 
\begin{equation}
\int dr\ dr'\ dt \ dt'\  \gamma_1(r-r')
{\tilde{\phi}}({\bf r},t){\tilde{\phi({\bf r}',t')}}
cos(\phi({\bf r},t)-\phi({\bf r}',t'))
(\nabla \phi({\bf r},t))^2.\label{gama1}
\end{equation}
It is possible to choose a symmetrized form for this vertex but 
the results will remain the same.

The nonlinearity also gets renormalized unlike the case with a local 
correlator of Ref \cite{tsai2}. The  diagrams contributing  to the 
renormalization of the nonlinearity is shown in Fig.3.
\vbox{
\begin{center}
\psfig{file=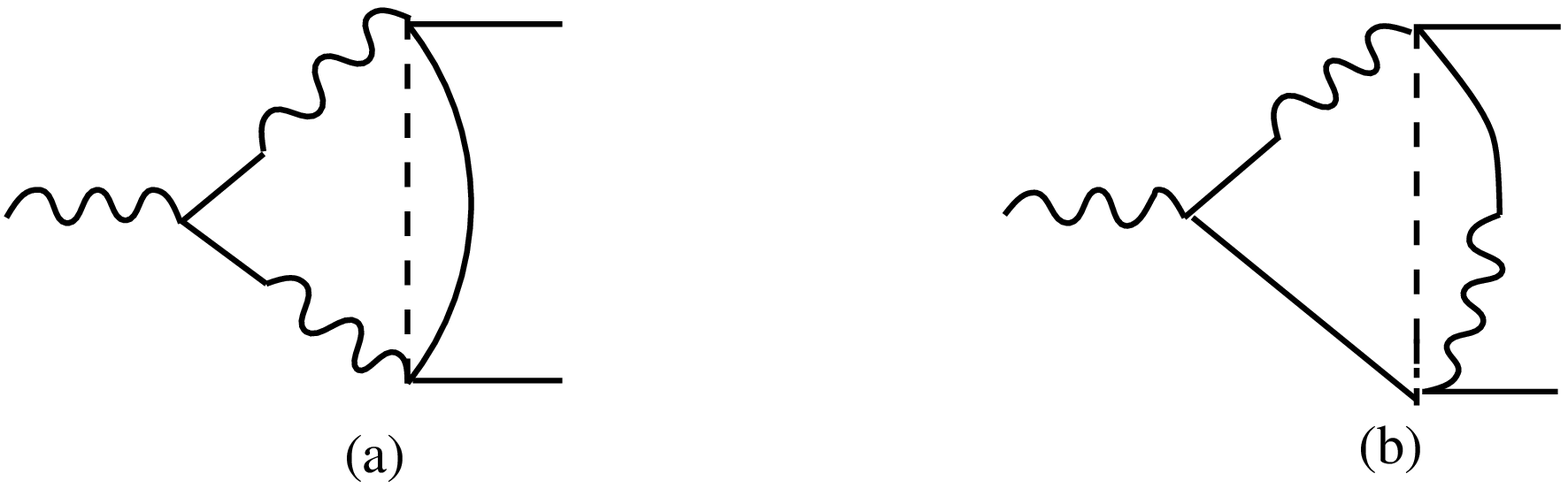,width=6in,angle=0}
{Fig. 3 Diagrams contributing to the renormalization of $\lambda$.}
\end{center}}
The details of the derivation of various terms present in the recursion
relations are described in the appendix. In appendix A,
 we show the necessary 
details to find out the renormalization of the tension. 
In appendix B and C, the renormalizations of the disorder 
correlator and the nonlinearity are discussed respectively. 
The final flow equations of the parameters in terms of a dimensionless 
quantity
$n_l=\frac{\mu \vartheta}{8 \pi \kappa}$ are given as
\begin{mathletters}
\label{recrel}
\begin{equation}
\frac{\partial \kappa}{\partial l}=
\frac{1}{8\pi \kappa}\int d{\bf r}A(r)\label{eq12a}
\end{equation}
\begin{equation}
\frac{\partial \mu^{-1}}{\partial l}=\frac{4 n_l}{\mu \vartheta}
\int dt \ d{\bf r}B(r,t)\label{eq12b}
\end{equation}
\begin{equation}
\frac{\partial \vartheta}{\partial l}= 4 n_l
\int d{\bf r} \ 
dt\ B(r,t) +
\frac
{\lambda^2 \mu \vartheta^2}{16 \kappa^3(2\pi)}\label{eq12c}
\end{equation}
\begin{equation}
\frac{\partial \gamma(r)}{\partial l}= (4+{\bf r}\cdot\nabla) \gamma(r)-2n_l
(\gamma(r)-
2\Lambda^2 \gamma_1(r))\label{eq12d}
\end{equation}
\begin{equation}
\frac{\partial \gamma_1(r)}{\partial l}=(2+{\bf r}\cdot\nabla)
\gamma_1(r)-\frac{n_l\lambda^2}
{64\kappa^2\Lambda^2}(\gamma(r)-4\gamma_1(r)\Lambda^2)\label{eq12e}
\end{equation}
\begin{equation}
\frac{\partial \lambda}{\partial l}=-\frac{\lambda}{16\pi\kappa^2} 
\int d{\bf r}A(r)\label{eq12f}
\end{equation}
\begin{equation}
\frac{\partial {\tilde{J}}}{\partial l}=
2 {\tilde {J}}+ n_l \lambda, \label{eq12g}
\end{equation}
\end{mathletters}
where $A(r)=\gamma(r) r^2 J_0(\Lambda r)e^{-Y(r,0)}$ and 
$B(r,t)=\gamma(r)J_0(\Lambda r)e^{-\mu\kappa\Lambda^2t}e^{-Y(r,t)}$.
The flow equation for $\gamma$ in this form apparently 
differs from that of ref. \cite{tsai2} for 
a short range correlation. They obtain a $\lambda$ dependent contribution
which in our formulation appears through a new vertex $\gamma_1$ generated
under renormalization. Since  $\gamma_1=0$ is the initial condition, one 
can  write for simplicity the flow of $\gamma$ as 
\begin{equation}
 \frac{\partial \gamma(r)}{\partial l}= (4+{\bf r}\cdot\nabla) \gamma(r)-2n_l
\gamma(r)-C \lambda^2 \gamma(r)\label{new},
\end{equation} where $C$ is a constant as it is
in ref. \cite{tsai2}. 

\section{Discussion}
For $\lambda=0$, the growth equation, Eq. \ref{eq2}, corresponds to the
equilibrium problem \cite{stef}.  In this case, we recover the equations 
of Ref. \cite{stef}. 
The recursion relation for the  correlator
of the form $\gamma(r) \sim c_0 r^{-2\alpha}$ for large $r$, implies 
that $\alpha$ 
remains unchanged under renormalization, as has been observed in ref. 
\cite{stef}. It is only the amplitude $c_0$ that is renormalized.
It is also apparent from Eq \ref{eq12d}, that 
for $\lambda=0$ 
the relevance or irrelevance of  $\gamma(r)$ is determined by the value 
of $n_l$, that plays the role of temperature. For a perfect crystal
with $\gamma(r)=g_0$, $n_l=2$
determines the roughening transition whereas 
for short range disorder, $\gamma(r)=g_0\delta(r)$, 
this boundary is at $n_l=1$\ \cite{tsai1,stef}.

For $\gamma(r)=g_0 \delta(r)$ 
the flow equation shows that $\lambda$
is invariant under renormalization. This is in 
agreement with ref \cite{tsai2}. For any other short range form of
$\gamma(r)$ (e.g. exponentially decaying $\gamma(r)$),
it is possible to see from Eq \ref{new}, that the width of the
correlation defined as $\int dr r^2 \gamma(r)/\int \gamma(r)dr$ vanishes 
as one approaches $l\rightarrow \infty$. Interestingly, for a long
range form of $\gamma(r)$ as mentioned above the integral converges
for $\alpha>1$. This $\alpha=1$ is the  borderline for the roughening and 
super roughening transition\cite{stef}. We therefore conclude that any 
short range
form of $\gamma(r)$ leads to the same conclusion as ref. \cite{tsai2},
namely that the large scale behavior of the growing surface is KPZ like and 
the super roughening transition is lost.
With a power law correlation of the disorder, 
this short range limit is achieved when $\alpha>1$. 

Let us now consider $0<\alpha<1$.
Note that the integral  over $r$
 in the flow equation for $\lambda$  is the same as
that in the flow equation for $\kappa$. Therefore with 
$\gamma(r)\sim c_0 r^{-2\alpha}$, this integral is positive
as long as $n>5/4-\alpha$ as in ref \cite{stef}. 
As a consequence of this  the nonlinearity 
decreases with the length scale. This is quite an 
astonishing feature arising due
to the nonlocality of the correlator. 
Recall that the $\lambda$ term occurs 
to take into account the lateral growth of 
the surface.  Our result suggests 
that at least at the
initial stage, the lateral growth of the evolving surface is inhibited
by the long distance correlation of the quenched substrate disorder.
The effect of this decrease of $\lambda$ implies that the super roughening 
transition is unaffected up to the length scale below 
which our treatment with a $F\rightarrow 0$ is valid.
If we stop at this leading order of the flow equations, then from
\cite{stef}, we can argue that for $\alpha<1$ one would observe 
roughening transition.
There is, however, a possibility of generation of a 
driving force due to the  nonlinearity $\lambda$ as given in 
the recursion relation \ref{eq12g}. 
The growth of this force with the length scale 
requires a non perturbative treatment of the 
force \cite{rost} (explained below) and the decay of the nonlinearity is 
eventually prevented.  The 
relevant length scale at which this crossover takes place, 
can be obtained by solving the coupled
differential equation with and without the force term.
Also note that, to this order, the flow of $\lambda$ and $\kappa$, under 
renormalization, has an invariance 
  $\lambda \kappa^{\frac{1}{2}}=constant$,
the physical origin of which is not clear to us.

Let us consider a finite force case. According to reference \cite{rost}
 a finite force for a perfect crystal ($\alpha = 0$) implies a rough 
surface with the 
height fluctuations having a power law scaling with the length. To treat 
a finite force we redefine the  height as 
\[\phi({\bf r},t)\rightarrow \phi({\bf r},t)+Ft\]
where $F$ is the external force. This implies a replacement of the disorder 
part of the action as
\begin{equation}
{\cal A}_0^{(d)}=\int dt dt' d{\bf r}d{\bf r'}
 \frac{1}{2}\gamma_0({\bf r}-{\bf r'})
{\tilde \phi}_0({\bf r},t){\tilde \phi}_0({\bf r'},t')
\cos[\phi_0({\bf r},t)-\phi_0({\bf r'},t')+(t-t')F]
\end{equation}
The next step is the 
splitting of cosine into  a product of sine and cosine terms.
Most crucial is the  renormalization of 
$\lambda$ which follows from the possible contraction of the  term
$\sin[\phi_0({\bf r},t)-
\phi_0({\bf r'},t')]\sin[(t-t')F]$. This leads to the  flow 
equation for $\lambda$ analogous 
to that in the reference \cite{rost}. We do not go 
into further details of the calculation since the
procedure for extracting 
this term is similar to the details given in the  appendices.  

The connection with reference \cite{krug} can also be understood in 
our approach. There the  equation of motion contains a random force. 
Generating functional obtained after averaging over this random 
force  is  in fact the first term of Eqn. \ref{eq8} in the expansion 
of the periodic cosine function. 
The random lateral drift velocity, as suggested in ref. \cite{krug},
appears through a new vertex $\gamma_1$ of Eq. \ref{gama1}
again as a first term in the expansion of the periodic function.
The random part of the KPZ nonlinearity is not present in our work 
but  it will obviously  be generated in higher order of $\lambda$ 
contributions.
In other words, the terms that have been discussed in Ref. \cite{krug} are 
naturally generated or present as the leading order terms 
in our functional renormalization scheme with 
a finite cut off, though they do not occur in the 
dimensional regularization scheme of Ref. \cite{tsai2}.
 We have treated the periodic function completely.
It is clear from section III that our  renormalization 
is crucially dependent on the periodic function, and in fact,
truncation to first order of the cosine will never lead to any
renormalization of $\lambda$.
 Similarly, the finite force scheme
is also dependent on the periodic function.  We are therefore not
sure whether the leading order terms in the expansion of the
cosine function of Eqs. \ref{eq8} and \ref{gama1} can capture
the whole effect, especially the 
question of a new universality class.  To address this question
one has to go to higher orders in RG.  This  remains to be done.

\section{Summary}
In this paper we investigated the super roughening transition in
presence of a nonlinearity supporting a lateral growth of the surface. 
The disorder is considered to have a long range correlation as 
discussed after Eq \ref{eq3}. A 
functional renormalization group analysis using a finite momentum cutoff 
leads to flow equations for the parameters of the growing surface 
described by Eq. \ref{eq2} and \ref{eq8}, and a flow equation for the 
disorder correlation.
These equations (\ref{eq12a}-\ref{eq12g})  predict the macroscopic 
properties of the surface. Our equations reduce to the known form of 
reference \cite{tsai2} for short range correlation of the disorder. 
The lateral growth governed  by the 
nonlinear term  with coefficient $\lambda$ is initially suppressed 
due to the long range disorder correlation. 
This is apparent from 
equation \ref{eq12f}. Therefore at least at the initial stage one 
might expect to see that the super roughening transition remains 
unaffected. Over a large length scale  a finite force
can ultimately lead to a rough surface and destroy the transition.
This requires a treatment of a finite force as 
discussed in section IV.  
In our analysis, we have taken a finite cutoff 
and have shown that the terms
that have been predicted from numerical lattice 
simulations are really 
generated though our analysis envisages a more general term.

\acknowledgments

I thank S. Scheidl for introducing me to this field. I also thank
T. Nattermann and S. M. Bhattacharjee for various useful comments.
Support from SFB 341 is acknowledged.
\appendix
\section{Renormalization of the  tension}
We discuss only the contribution of the pinning potential $\gamma(r)$ to the 
renormalization of the tension $\kappa$. The renormalization of the other 
quantities like temperature, and mobility can be found out in an analogous
way with appropriate diagrams.
One representative Feynmann diagram that contributes to the 
renormalization of the  tension is the shown in Fig 4.
\vbox{\begin{center}
\psfig{file=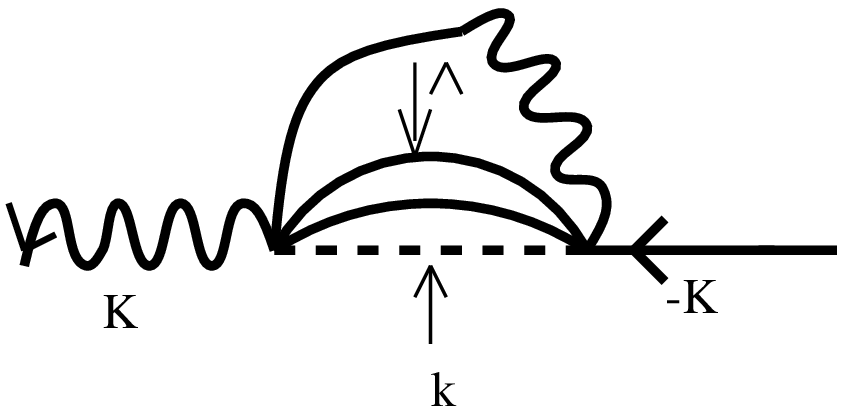,width=1.5in,angle=0}
{Fig. 4 Diagram contributing to the renormalization of the 
tension $\kappa$.}
\end{center}}       
The external lines are  associated with   momentum ${\bf K}$ and all 
the internal lines contain fast momenta. However the sum of all 
of the internal momenta should add up to yield $K$. Let us choose 
one internal $C$ line with momentum $\Lambda$ as shown in Fig 4. Its 
contribution in real space is 
\begin{equation}
 \int d{\bf p} \ \Lambda \delta(p - \Lambda) 
\exp(i {\bf p}\cdot {\bf r}) C(p,t)   
= \frac{\Lambda^2}{2\pi} J_0(\Lambda r) \frac{\mu \vartheta}{2\kappa\Lambda^2}
 \exp 
(-\mu \kappa \Lambda^2\mid t \mid ) \equiv G^{\phi,\phi}(r,t). 
\end{equation}
Therefore the contribution of Fig4 is
\begin{equation}
 \frac{1}{2}\int dt\ \int d{\bf r} \int d{\bf k}  
\exp [-i ({\bf K} - {\bf k})\cdot 
{\bf r}] G^{\phi\phi}(r,t) R(r,t) \exp[-Y(r,t)]\gamma(k)\label{eqap2}
\end{equation}
where $\exp[-Y(r,t)]$ takes care of the sum over all possible 
diagrams that arise from all possible contractions of the internal lines 
as it is in the case of local correlator \cite{tsai1,tsai2}. 
The other possibility arises
when one associates the $R$ line with the momentum $\Lambda$.
In this case we have the line with the fast momentum as
\begin{equation}
\int d{\bf p} \ \Lambda \delta(p - \Lambda) 
\exp(i {\bf p}\cdot {\bf r}) R_0(p,t)
=\frac{\Lambda^2}{2\pi} J_0(\Lambda r)\mu e^{-\mu \kappa \Lambda^2 t}
\equiv G^{\phi,{\tilde \phi}}(r,t),
\end{equation}
and the full expression is
\begin{equation}
\frac{1}{2}\int dt\ \int d{\bf r} \int d{\bf k}  \exp [-i ({\bf K} -
 {\bf k})\cdot {\bf r}]
G^{\phi{\tilde \phi}}(r,t) \exp[-Y(r,t)]\gamma(k)\label{eqap1}
\end{equation}
The flow equation for the tension is obtained by 
addition of 
Eqs. \ref{eqap1} and \ref{eqap2} with an extra symmetry 
factor, completing the integral over ${\bf k}$ and extracting the term 
proportional to $K^2$.
Simplified form given in Eq. \ref{eq12a} is achieved by the use of the 
FDT and a subsequent integration by parts over $t$.

\section{Renormalization of $\gamma$}

The first term in the flow equation of $\gamma$ is, in fact, the result of 
the contribution that comes from expanding the cosine in  the 
pinning potential. The simplest diagram in the one loop level is 
shown in Fig 5.
\vbox{\begin{center}
\psfig{file=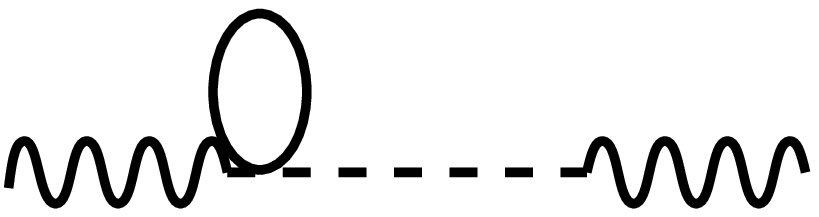,width=1.5in,angle=0}
{Fig. 5 Diagram contributing to the renormalization of $\gamma$}
\end{center}}
The corresponding expression is 
\begin{equation}
\frac{\gamma(r)}{2}\ \int \frac{d{\bf k}}{(2\pi)^2} \Lambda \delta(k-\Lambda) C(k,0).
\end{equation}
This leads to the first term of the  RG contribution in Eq. \ref{eq12d}.
The second term with $\gamma_1$ in Eq. \ref{eq12d} will be clear from 
the following discussion.
The contribution that comes from the combined effect of the nonlinearity 
is at the two loop level and the possible diagrams are shown explicitly
in Ref \cite{tsai2}.
 The effect of these diagrams are incorporated through a higher 
order vertex. Usual procedure that is adopted is to snip off the internal 
line \cite{shankar}
and generate a new vertex. This vertex will renormalize  the vertex $\gamma$
when they are rejoined again. We pick up one such two loop 
diagram which contributes in the field theoretic approach. Snipping off 
one internal line leads to a vertex of the form given in Eq. \ref{gama1}.

We call this new vertex $\gamma_1(r)$ which differs from the 
$\gamma(r)$ vertex by
 two $\nabla \phi$ type legs. Once  they are rejoined 
this vertex will contribute to renormalization of $\gamma(r)$. The second 
term with $\gamma_1$ in Eq. \ref{eq12d} follows from this prescription.
We now discuss the renormalization of  the vertex $\gamma_1$.
There are two such possible diagrams which can originate from the 
original two loop diagrams and lead to the renormalization of $\gamma_1$.
From the other two loop diagrams it is easy to observe that no such
 vertex that respects the symmetry of the system can be formed. 
The contribution of the $\gamma_1$ vertex in the renormalization of 
$\gamma$ is 
identical to the contribution of $\gamma$ vertex itself except 
for a multiplicative
 factor $\Lambda^2$ which is obvious from a dimensional analysis.
As it is evident from Fig 2, this new vertex is generated from the term of 
$O(\lambda^2 \gamma)$. 
It is easy to evaluate the expressions of the diagrams in Fig2.
In the momentum and frequency representation,
\begin{eqnarray}
&& Fig2b=
-\frac{\lambda^2}{4}\gamma(r)\int_0^{\infty}dt_2 \int_0^{t_2} dt_1
\int dk_2  R(k_{2m1},t_2-t_1)
C(k_{2m1},t_2)
R(k_{2p1},t_1)
\nonumber\\ &&\ \ \   
({\bf k}_1\cdot {\bf k}_{2p1})({\bf k}_1\cdot {\bf k}_{2m1})\nonumber\\
&&=-\frac{\mu\vartheta}{8\pi\kappa} \frac{\lambda^2}
{16\kappa^2\Lambda^2}\frac{\gamma(r)}{4} k_1^2,
\label{eq14}
\end{eqnarray}
where ${\bf k}_{2p1} =  {\bf k}_2+{\bf k}_1/2$, and
${\bf k}_{2m1} =  {\bf k}_2-{\bf k}_1/2$
in the hydrodynamic limit i.e $k_1\rightarrow 0$.
Similarly
\begin{equation}
Fig2c=\frac{\mu\vartheta}{8\pi\kappa} 
\frac{\lambda^2}{8\kappa^2\Lambda^2} \frac{\gamma(r)}{4}k_1^2.
\label{eq15}
\end{equation}
A similar contribution comes from the vertex involving $\gamma_1$. The 
combined expression leads to the recursion relation for $\gamma_1$ in Eq. 
\ref{eq12e}. The new vertex 
$\gamma_1$ is completely the effect of renormalization. 

\section{renormalization of $\lambda$}
We notice that the possible one loop diagrams which contribute to the 
renormalization of the nonlinearity $\lambda$ in the case of nonlocal
correlator are those shown in Fig 3.
 There are other possible diagrams 
that could lead to a contribution in the renormalization of the nonlinearity. 
But we note that they cancel altogether. For example the diagrams shown 
in Fig 6. 
The expression corresponding to the diagrams in Figure 3a is
\begin{equation}
Fig3a=-\frac{\lambda}{4}
\int \frac{d{\bf k}_1}{(2\pi)^2} d\omega_1 R(k_1,\omega_1)R(k_1,-\omega_1)
\int d{\bf r}\int_0^{\infty} dt \gamma(r)\ \frac{k_1^2 r^2}{2} e^{-Y(r,t)}
e^{i{\bf k}_1\cdot {\bf r}}e^{-i\omega_1 t}.\label{eq16}
\end{equation}
After the integration over the shell  momentum $k_1$
we obtain
\[{\rm Fig  3a}
=-\frac{\mu^2\lambda \Lambda^4}{32 \pi^2}
\int d\omega_1 \frac {1}{(\mu\kappa \Lambda^2)^2+\omega_1^4}
\int_0^{\infty} dt
 \int d{\bf r} \  \gamma(r)\ r^2 J_0(\Lambda r)e^{\-i\omega_1 t}e^{-Y(r,t)}
\]
Furthermore the  integration over the 
frequency leads to the final contribution of Fig 3a as
\begin{equation}
-\frac{\mu\lambda\Lambda^2}{32\pi \kappa}\int_0^{\infty} dt\ d{\bf r}\ 
 r^2\ \gamma(r)\ 
J_0(\Lambda r)e^{-Y(r,t)}e^{-\mu\kappa\Lambda^2 t}.\label{eq17}
\end{equation} 
Similarly the expression corresponding to Fig 3b is 
\begin{equation}
Fig 3b=-\frac{\lambda}{16}
\int \frac{d{\bf k}_1}{(2\pi)^2} \ d\omega_1 R(k_1,\omega_1)C(k_1,\omega_1)
\int d{\bf r} \int_0^{\infty}dt r^2 R(r,t) \gamma(r) k_1^2 r^2
e^{i\omega_1 t} e^{-Y(r,t)}.\label{eq18}
\end{equation}
After performing the integration over the momentum shell and the frequency,
we get,
\begin{equation}
{\rm Fig \ 3b} =-\frac{\lambda\vartheta\mu}{128\pi\kappa^2}\int d{\bf r} 
\int _0^{\infty} dt
r^2 \gamma(r) R(r,t)J_0(\Lambda r)e^{-\mu\kappa \Lambda^2 t}
e^{-Y(r,t)}\label{eq19}
\end{equation}
An use of the relation
\begin{equation}
\theta(t>0)\partial_t e^{-Y(r,t)}=-\frac{\mu\vartheta}{2} R(r,t)e^{-Y(r,t)},
\label{eq19a}
\end{equation} and subsequent integration over $t$ 
leads to the following expression
\[{\rm Fig \ 3b}=-\frac{\lambda}{64\pi\kappa^2} 
\int d{\bf r} r^2 J_0(\Lambda r)\gamma(r)e^{-Y(r,0)}\]
\begin{equation}
+\frac{\lambda\mu\Lambda^2}{64\pi\kappa}\int d{\bf r} 
\int_0^\infty dt r^2 J_0
(\Lambda r) \gamma(r)e^{-Y(r,t)} e^{-\mu\kappa\Lambda^2 t}.\label{eq19B}
\end{equation}
Combining the contributions of Fig 3a and Fig 3b with appropriate symmetry
factor we arrive at the recursion relation for $\lambda$.

The diagrams corresponding to Fig 6 can be evaluated in a similar manner.
\vbox{
\begin{center}
\psfig{file=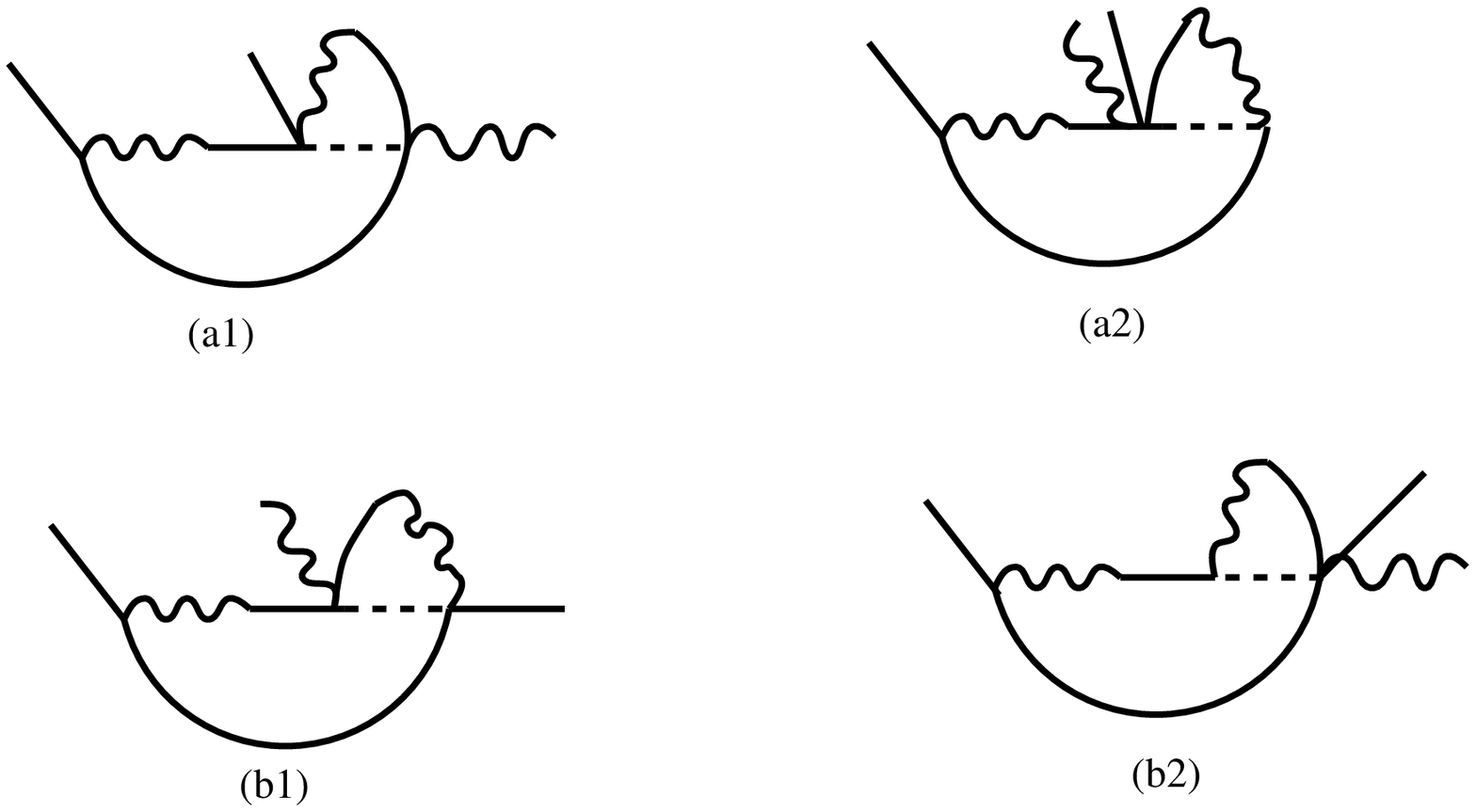,width=6in,angle=0}
{Fig. 6 Diagrams which could  contribute to the renormalization 
of $\lambda$ but vanish due to their mutual cancellation.}
\end{center}}
\begin{eqnarray}
Fig 6a1=
\int \frac{d{\bf k}}{(2\pi)^2}\int d\omega R(p_+,\omega)C(p_-,\omega) 
{\bf p}\cdot {\bf p}_-\nonumber\\
\int d{\bf r}
 \int_0^{\infty} dt \gamma(r)R(r,t)J_0(p_-r)e^{-Y(r,t)}
e^{i\omega t},\label{eq20}
\end{eqnarray}
\begin{eqnarray}
Fig6b1=\int \frac{d{\bf k}}{(2\pi)^2}
\int d\omega R(p_+,\omega)C(p_-,\omega) {\bf p}
\cdot {\bf p}_-\nonumber\\
\int d{\bf r} \int_0^{\infty} dt \gamma(r) R(r,t) e^{-i\omega t} e^{-Y(r,t)} 
J_0(p_+ r)
\label{eq21}
\end{eqnarray}
where ${\bf p}_{+} = {\bf p}/2 + {\bf k}$, and 
${\bf p}_{-} = {\bf p}/2 - {\bf k}$. 
In the above expressions ${\bf p}$ is the slow momentum and the contribution 
to the $\lambda$ vertex can arise from the following terms after 
adding Fig 6a1 and 6b1
\begin{eqnarray}
&&
\int d{\bf r} \int_0^{\infty} dt\ \gamma(r)J_0(\Lambda r)\ R(r,t) e^{-Y(r,t)}\ 
\int \frac{d{\bf k}}{(2\pi)^2}
2p_+ ^2 \ {\bf p}\cdot{\bf (p/2-k)}\nonumber\\&&
 \int d\omega \ e^{i\omega t}
\frac{1}{(p_+^4+\omega^2)(p_-^4+\omega^2)}\nonumber\\
&&+\int d{\bf r} \int_0^{\infty} dt 
\ r\ \gamma(r) J_0(\Lambda r)\ R(r,t) e^{-Y(r,t)}
 \int \frac{d{\bf k}}{(2\pi)^2} {\bf p}\cdot {\bf (p/2-k)} \nonumber\\&&
\int d\omega e^{i\omega t} 
\frac{{\bf p}\cdot {\bf k}\ (-2 i \omega)}{(p_+^4+
\omega^2)(p_-^4+\omega^2)}\label{eqn22}
\end{eqnarray}
We need not go into the final expression corresponding to the combined 
contribution of Fig 6a1 and Fig 6b1. 
This is because, interestingly, the  other two diagrams in Fig 6a2 and 6b2 
lead to the
same contribution but {\it with opposite sign}, thereby 
canceling the contribution of Figs 6a1 and 6b1.  This cancellation
leads to a considerable simplification of the recursion relations.


\begin{references}
\bibitem{van} H. van Beijeren and I. Nolden in {\it Structure and Dynamics
of Surfaces II}, edited by W. Schommers and P. von Blanckenhagen 
(Springer, Berlin, 1987)
\bibitem{noz} P. Noziers and F. Gallet, J. Phys. (Paris) {\bf 48}, 353 (1987),
P. Noziers, in {\it Solids far from equilibrium}, edited by C. Godreche (
Cambridge University Press, Cambridge, England, 1992)
\bibitem{chui} S. T Chui and J. D. Weeks, Phys. Rev. Lett. {\bf 38}, 
4978 (1976)
\bibitem{tsai1} Y. -C. Tsai and Y. Shapir, Phys. Rev. Lett. {\bf 69}, 1773
(1992), Phys. Rev. E {\bf 50}, 3546 (1994)
\bibitem{cardy}J. J. Cardy and S. Ostlund, Phys. Rev. B {\bf 25}, 6899 
(1982); J. Toner and D. P. Divincenzo, Phys. Rev. B {\bf 40}, 632 (1990)
\bibitem{korsh} S. E. Korshunov, Phys. Rev. B {\bf 48}, 3969(1993);
T. Giamarchi and P. LeDoussal, Phys. Rev. Lett. {\bf 72}, 1530 (1994)
\bibitem{cule} D. Cule and Y. Shapir, Phys. Rev. Lett. {\bf 74}, 114 (1995),
H. Rieger, Phys. Rev. Lett. {\bf 74}, 4964 (1995)
\bibitem{shapir} C. Zeng, A. A. Middleton and  Y. Shapir, Phys. Rev.
Lett. {\bf 77}, 3204 (1996)
 Condmat/9609029
\bibitem{stef} S. Scheidl, Phys. Rev. Lett. {\bf 75}, 4760 (1995)
\bibitem{kpz}M. Kardar, G. Parisi, and Y. C. Zhang, Phys. Rev. Lett. 
{\bf 56}, 889 (1986),  E. Medina, T. Hwa, M. Kardar, and Y. C. Zhang,
Phys. Rev. A {\bf 39}, 3053(1989)
\bibitem{rost} M. Rost and H. Sphon, Phys. Rev. E {\bf 49}, 3709 (1994)
\bibitem{tsai2} Y. -C. Tsai and Y. Shapir, Phys. Rev. E {\bf 50}, 
4445 (1994)
\bibitem{krug} J. Krug, Phys. Rev. Lett. {\bf 75}, 1795 (1995)
\bibitem{halpin} For a recent review, see T. Halpin-Healy and  Y-C Zhang,
Physics Reports, {\bf 254}, 215 (1995)
\bibitem{natter} T. Nattermann, S. Stepanow, L-H. Tang and 
H. Leschhorn, J. Phys. (France) {\bf 2},1483 (1992)
\bibitem{martin} P. C. Martin, E. D. Siggia and H. A. Rose,
 Phys. Rev. A {\bf 8}, 423 (1973)
\bibitem{comm} In this formalism, the generating function describes
the probability of the height variable $\phi$. We therefore, can average 
the generating function over the disorder.
\bibitem{culey} D. Cule and Y. Shapir, Phys. Rev. B {\bf 51}, R3305 (1995)
\bibitem{shukla} P. Shukla in {\it {Models and techniques of Statistical 
Physics}}, edited by S. M. Bhattacharjee, (to be pubished by Narosa, 1996)
\bibitem{wegner} F. J. Wegner and A. Houghton, Phys. Rev. A {\bf 8}, 401
 (1973)
\bibitem{gs} Y. Y. Goldschmidt and B. Schaub, Nucl. Phys. {\bf B251},
 77, (1985)
\bibitem{shankar} See e.g., R. Shankar, Rev. Mod. Phys. {\bf 66} 129 (1994)
\end{references}
\end{document}